\documentclass[american,aps,pra,reprint,superscriptaddress,twocolumn]{revtex4-1}
\usepackage[colorlinks,citecolor=red]{hyperref}
\hypersetup{colorlinks,linkcolor=myurlcolor,citecolor=myurlcolor,urlcolor=myurlcolor}
\usepackage{braket,colortbl,amsthm,amsmath,amssymb,cleveref,txfonts}
\definecolor{myurlcolor}{rgb}{0,0,0.7}
\usepackage{graphics,graphicx}
\usepackage{color}

\theoremstyle{plain}

\def\bea{\begin{eqnarray}}
\def\eea{\end{eqnarray}}
\def\ba{\begin{array}}
\def\ea{\end{array}}

\def\ket{\rangle}
\def\bra{\langle}
\def\beq{\begin{equation}}
\def\eeq{\end{equation}}

\begin{document}
\title{Generating and detecting bound entanglement in two-qutrits using a family of indecomposable positive maps}

\author{Bihalan Bhattacharya}
\email{bihalan@gmail.com}
\author{Suchetana Goswami}
\email{suchetana.goswami@gmail.com}
\affiliation{S. N. Bose National Centre for Basic Sciences, Block JD, Sector III, Salt Lake, Kolkata 700 106, India}
\author{Rounak Mundra}
\affiliation{Center for Security Theory and Algorithmic Research,
International Institute of Information Technology, Gachibowli, Hyderabad, India}
\author{Nirman Ganguly} 
\affiliation{Department of Mathematics, Birla Institute of Technology and Science Pilani, Hyderabad Campus, Telangana-500078, India.}
\author{Indranil Chakrabarty}
\affiliation{Center for Security Theory and Algorithmic Research,
International Institute of Information Technology, Gachibowli, Hyderabad, India}
\author{Samyadeb Bhattacharya}
\email{samyadeb.b@iiit.ac.in}
\affiliation{Center for Security Theory and Algorithmic Research,
International Institute of Information Technology, Gachibowli, Hyderabad, India}

\author{A. S. Majumdar}
\affiliation{S. N. Bose National Centre for Basic Sciences, Block JD, Sector III, Salt Lake, Kolkata 700 106, India}

\begin{abstract}
The problem of bound entanglement detection is a challenging aspect of quantum information theory for higher dimensional systems. Here, we propose an indecomposable positive map for two-qutrit systems, which is shown to detect a class of positive partial transposed (PPT) states. A corresponding witness operator is constructed and shown to be weakly optimal and locally implementable. Further, we perform a structural physical approximation of the indecomposable map to make it a completely positive one, and find a new PPT-entangled state which is not detectable by certain other well-known entanglement detection criteria. 
\end{abstract}

\maketitle

\section{Introduction}

The inseparable feature of quantum states \cite{EPR_35, S_35, R_89, B_64} plays the most crucial role in various information processing tasks \cite{BW_92, BBCJPW_93, BCWSW_12, AMP_12}. Entanglement is the central feature of the theory of quantum information science and the detection of entanglement in an arbitrary quantum system is considered to be one of the most fundamental aspects of the subject. The most effective way to detect entanglement theoretically, is via the usage of positive but not completely positive (NCP) maps, of which the most famous and heavily utilized example is given by the partial transposition (PT) map \cite{pt1}. 

It is well known that PT gives us a necessary and sufficient criterion, named the separability criterion to detect entanglement only for $2\times 2$ and $2\times 3$ dimensional states \cite{ppt1}. It is seen that for these dimensions, all  entangled states have negative partial transposition (NPT). There are different prescribed protocols for detection of two-qubit entanglement based on this criterion \cite{ADH_08, GA_12, ZGZG_08, CHKST_16, GCGM_19}. On the other hand, entanglement detection in general is a NP hard problem \cite{posit8}. In case of higher dimensional systems, there exists a class of states which are entangled but having a positive partial transposition (PPT), and hence cannot be detected by the NPT criterion.

The entanglement of PPT entangled states is not distillable \cite{bound1}. The presence
of bound entanglement in such states has evoked much interest as to the possibilities
of using or unlocking the entanglement in present in them \cite{bound2, bound3}. 
Bipartite bound entanglement channels can exhibit superadditivity of quantum channel
capacity \cite{bound4}. A further interesting and difficult task is to detect such
bound entanglement \cite{bound5, bound6}, and methods have been recently suggested
to prepare and certify bound entangled states that are robust for experimental
verification \cite{bound7}. The bound entanglement in PPT entangled states is inextricably linked to indecomposable positive maps.

The structure of positive maps has been an area of interest to mathematicians for a long period of time, since it is extremely hard to determine the positivity of a map even in low dimensions. Ever since the seminal works of Peres and Horodecki \citep{Peres96, pt1}, it has been clear that such maps play an instrumental role in detection of quantum entanglement. Considerable effort from both mathematicians and physicists \citep{Stinespring55, arveson69, Stormer82, Worono76,  Tomiyama85,  Osaka91,  Cho92, Ha03, Majewski01, Kossakowski03, Piani06, Majewski07, Sarbicki12, Zwolak13, Marciniak13,  Miller15, Rutkowski15, Lewenstein16, Marciniak17} have shed some light on the structural intricacies of positive maps and their applications in physics. Applications of positive maps in the study of entanglement theory have catalysed  the development of both domains.

Indecomposable positive maps play a key role in generating entangled states in higher dimensions. The class of positive maps which can be decomposed as an algebraic sum of two relatively simple convex sub classes of positive maps, {\it viz}., completely positive class and completely co-positive class is called decomposable. Since transposition maps are completely co-positive in nature, quantum states having PPT can not be detected by them. As a consequence, indecomposable maps are important for detecting PPT entangled states.  Therefore, constructing  non completely positive maps for detecting PPT entangled states is of considerable importance in  entanglement theory.

As the PPT criterion fails to detect bound entanglement in higher dimensions,  certain other criteria have been proposed in the literature which can detect some PPT entangled states. These include the computable cross norm or realignment criterion (CCNR criterion) \cite{R_03, CW_02}, range criterion \cite{BDMSST_99, BP_00}, covariance matrix criterion (CMC) \cite{GHGE_07, GGHE_08} and others.  In the present work we further explore the connection between the theory of positive maps and entanglement. We introduce an indecomposable positive map on the algebra of $3 \times 3$ complex matrices to obtain a PPT entangled state of a two-qutrit system.  
Our proposed non-completely positive map not only detects a class of two-qutrit bound entangled states, but also introduces a class of them which are not detected by several of the previously mentioned criteria. 

Since non-complete positive maps correspond to unphysical operations, it is impossible to implement them in the laboratory. However, it is indeed possible to construct a physically implementable complete positive map from a given unphysical map using the notion of structural physical approximation (SPA) \cite{spa1, spa2} which we employ in this work.
The SPA technique has also been used for realization of the optimal singlet fraction \cite{satya}. On the other hand, PPT entangled states have been constructed earlier from indecomposable positive maps  \cite{Ha03, HA04}.  Constructions of such states were done by exploiting the facial structures and various duality relations of the cone of positive maps. Here we devise a 
different method of contructing PPT entangled states via  usage of the structural physical approximation (SPA) \cite{spa1, spa2}. 

Three-level systems are of primary importance in laser physics, and possess features of interest from the quantum information perspective, as well \cite{qut1, qut2, qut3, qut4}. In practical quantum information procesing, detecting entanglement of a given unknown system and its quantification is one of the important areas of research. The theory of entanglement witnesses \cite{wit1, wit3, wit4, wit2, guehne, Ganguly09} provides a useful avenue to this end, and futher helps to identify resources useful for various information processing tasks \citep{Ganguly11, Adhikari12, Ganguly14, patro, vempati}.  Here we formulate a weakly optimal \cite{badzi} indecomposable entanglement witness from the positive map of our construction. This entanglement witness is shown to detect the proposed two-qutrit bound entangled state, and is further shown to be implementable through local operations. 

The structure of the paper is the following. In Section II, we discuss some prerequisites of the theory applied in the later sections. In Section III, we define a new one parameter family of indecomposable positive maps and show that it can detect a class of two-qutrit entangled states. In Section IV we  construct a weak optimal witness, which for a particular choice of parameter,  detects at least one class of bound entangled states. In Section V, we employ the structural physical approximation to construct a new class of PPT entangled states. We conclude in Section VI  with a summary of our results.

\section{Preliminaries}

In this section we shall discuss some preliminary details of positive maps. One can find detailed discussions on positive maps in \cite{book1, book2}. We consider Hilbert space of finite dimension, and shall deal with positive maps between algebra of matrices. 
The seminal results by Stormer \citep{stormer63} and Woronowicz \citep{Worono} showed that if $\mathcal{H}_1$
and $\mathcal{H}_2$ be two Hilbert spaces, then all positive maps acting on the set of bounded operators on $\mathcal{H}_1$ into  the set of all bounded operators acting on $\mathcal{H}_2$ are decomposable if product of the dimension of $\mathcal{H}_1$ and $\mathcal{H}_2$ is upper bounded by 6. The first example of indecomposable map was provided by M.D.Choi \citep{Choi75}, popularly known as Choi map. A new family of indecomposable map was considered by Hall \citep{Hall2006} and Bruer \citep{Bruer2006}. Later this map was generalised to a class of positive maps by Chruchinski and Kossakowski \citep{Chruchinski2008} and discussed the indecomposability and atomicity of the part of the class.  On the other hand, as discussed earlier, the theory of positive maps has a deep connection with quantum inseparability which instigates a new insight into the subject \citep{Peres96, pt1, PH1997, Terhal2002}.  

Here, we concentrate on the bipartite scenarios and recapitulate a few notions on separability and positive maps form the literature. As mentioned in the previous paragraph, if a bipartite state ($\rho_{AB}= \sum_{ijkl} p_{kl}^{ij} |i\ket \bra j| \otimes |k \ket \bra l| $) is a separable one, then it is PPT \cite{P_96, HHH_01}, where the partial transposition (with respect to the second subsystem) is given by, $\rho^{T_{B}}=\sum_{ijkl} p_{lk}^{ij} |i \ket \bra j| \otimes |k \ket \bra l| $. 
 In this case, a state $\rho$ can be concluded as a separable one if and only if for any positive map $\Lambda$, we have $(\openone \otimes \Lambda) \rho \geq 0$. Though there are a few examples of such maps \cite{choi,posit1,posit2,posit3,posit4,posit5,posit6,posit7} which can detect PPT entanglement, they are far from exhaustive.

Let $\mathbb{C}^{d}$  be the  complex Hilbert spaces of dimension $d$. 
Let $\mathcal{B} \left(\mathbb{C}^{d} \right)$  denote the space of all operators acting on $\mathbb{C}^{d}$.   $\mathcal{B} \left( \mathbb{C}^{d} \right) $ is endowed with Hilbert-Schmidt inner product defined by $<X, Y>= Tr \left[ X^{\dagger} Y \right] $ for any two members $X , Y \in \mathcal{B} \left( \mathbb{C}^{d} \right) $. The sub collection of $\mathcal{B} \left( \mathbb{C}^{d} \right)  $  consisting of hermitian, positive semidefinite operators having unit trace is known as the set of density operators acting on $\mathbb{C}^{d} $. 

Recall that operators acting on  finite dimensional spaces are bounded and can be represented as matrices with respect to some basis. Let $ \mathbb{M}_{d}$ and  $ \mathbb{M}_{k}$ be the algebra of $d \times d$ and $k \times k$ matrices respectively, over the field of complex numbers. A linear map $\Lambda : \mathbb{M}_{d} \rightarrow \mathbb{M}_{d} $ is said to be positive if $\Lambda \left( X \right) \geq \Theta $ for any positive semi-definite $X \in \mathbb{M}_{d} $, where $\Theta$ denotes the zero operator. A linear map is said to be k-positive if the map $\mathbb{I}_k \otimes \Lambda : \mathbb{M}_{k} \otimes \mathbb{M}_{d} \rightarrow \mathbb{M}_{k} \otimes \mathbb{M}_{d} $ is positive for some $k \in \mathbb{N}$. A linear map is said to be completely positive if it is k-positive for all $k \in \mathbb{N}$. Similarly a linear map $\Lambda$ is said to be k-copositive if $\mathbb{I}_k \otimes \left( \Lambda  \circ T \right) $ is positive for some $k \in \mathbb{N}$ and completely co-positive if $\Lambda  \circ T$ is completely positive, where $T$ stands for the transposition map.

Given any linear map $\Lambda : \mathbb{M}_{d} \rightarrow \mathbb{M}_{d}$, in connection with the celebrated Choi-Jamiolkowski isomorphism we can construct a matrix $\mathcal{C}_{\Lambda}$, known as Choi matrix, living in $\mathbb{M}_{d} \otimes \mathbb{M}_{d}$. Choi matrix can be obtained via the rule, $\mathcal{C}_{\Lambda} = \mathbb{I} \otimes \Lambda \left( \vert \phi^{+} \ket \bra \phi^{+} \vert  \right) $, where $\vert \phi^{+} \ket = \frac{1}{\sqrt{d}} \sum_{i=0}^{d-1} \vert ii \ket$ is the maximally entangled state in $\mathbb{C}^{d} \otimes \mathbb{C}^{d}$ and $\left\lbrace \vert i \ket _{0}^{d-1} \right\rbrace  $ stands for standard computational basis for $\mathbb{C}^{d}$. A linear map $\Lambda$ is completely positive iff its Choi matrix  $\mathcal{C}_{\Lambda}$ is positive semi-definite. It is to be noted that if a linear map is positive but not completely positive, then there exists some density operator $\rho$ whose image is not positive. Such an operator $\rho$ can not be a separable one. Hence,  positive but not completely positive maps can be used to detect entangled density operators. 

Another important notion of positive maps is their decomposability. A positive map $\Lambda$ is known to be decomposable if it can be expressed as $\Lambda = \Lambda_1 + \Lambda_2 \circ T$ where $\Lambda_1$ and $\Lambda_2$ are completely positive maps and $T$ denotes the action of transposition. Otherwise, it is said to be indecomposable. It is to be noted that decomposable maps can not detect PPT entangled density operators. Recall that a density operator $\sigma$ is said to be PPT if $ \left(\mathbb{I} \otimes T \right) \sigma  \geq \Theta$ . Therefore, indecomposable maps must detect at least one PPT entangled density operator. Moreover, a positive linear map is called atomic if it can not be expressed as a sum of 2-positive and 2-copositive map. An atomic linear map is by definition indecomposable.\\

Additionally, a linear map is said to be trace preserving if $ Tr \left[ \Lambda \left( X\right)  \right]= Tr\left[ X \right] ~~\forall X \in  \mathbb{M}_{d}$.  A linear map is said to be hermiticity preserving if $\Lambda \left( X \right)^{\dagger} = \Lambda \left( X^{\dagger} \right) ~~ \forall X \in  \mathbb{M}_{d} $.  Given a linear map $\Lambda : \mathbb{M}_{d} \rightarrow \mathbb{M}_{d}$, its dual map $\Lambda^{\dagger} : \mathbb{M}_{d} \rightarrow \mathbb{M}_{d}$ is defined by the relation $<\Lambda^{\dagger} \left( X \right), Y>  =  <X, \Lambda \left( Y \right)> $ for any operator $X, Y \in \mathbb{M}_{d}$. A map $\Lambda$ is positive iff its dual map $\Lambda^{\dagger}$ is also positive.
Using the above properties of  positive maps, in the next section we shall introduce a new class of  indecomposable positive maps. 

\section{One parameter family of indecomposable positive maps}

We now introduce a one parameter family of positive maps containing a clear indecomposable  subfamily. For this purpose, we start with the following definition.

\textbf{Definition 1:} Let $\mathbb{M}_3$ denote the algebra of $3 \times 3$ matrices over the field of complex numbers $\mathbb{C}$. We define a one parameter class of linear trace preserving maps $\Lambda_{\alpha} : \mathbb{M}_3 \rightarrow \mathbb{M}_3 $ by,
\begin{eqnarray}
\Lambda_{\alpha} \left( X \right) = \frac{1}{\alpha + \frac{1}{\alpha}} \begin{bmatrix}
\alpha (x_{11}+ x_{22})& -x_{12}&- \alpha x_{13}\\
-x_{21}&\frac{x_{22}+x_{33}}{\alpha}&-x_{32}\\
- \alpha x_{31}&-x_{23}&\alpha x_{33}+ \frac{x_{11}}{\alpha}
\end{bmatrix} 
\end{eqnarray}
\begin{eqnarray}
where~ X= \begin{bmatrix}
x_{11}&x_{12}&x_{13}\\
x_{21}&x_{22}&x_{23}\\
x_{31}&x_{32}&x_{33}&
\end{bmatrix} \in \mathbb{M}_3 ~~and ~~\alpha \in ( 0 , 1 ].
\end{eqnarray}


\noindent \textbf{Theorem 1:} ~~$\Lambda_{\alpha}$ is a positive map on $\mathbb{M}_3$ for all $0 < \alpha\leq 1$.\\\\
\noindent \textbf{Proof:}  To prove that the linear map is positive, it is sufficient to show that if acted upon any arbitrary pure state $|\phi\ket =(\phi_1,\phi_2,\phi_3)^T$, the map will give only positive semidefinite output. Here $\phi_1, \phi_2,\phi_3$ are arbitrary complex numbers with the constraint $|\phi_1|^2+|\phi_2|^2+|\phi_3|^2=1$. 

Here we have 
\begin{eqnarray}
\begin{array}{ll}
\Lambda\left(|\phi\ket\bra\phi|\right)=  \\ \frac{1}{\alpha + \frac{1}{\alpha}}\begin{bmatrix}
\alpha (|\phi_1|^2+ |\phi_2|^2) & -\phi_1\phi_2^{*} &- \alpha\phi_1\phi_3^{*}\\
 -\phi_1^*\phi_2&\frac{|\phi_2|^2+|\phi_3|^2}{\alpha}&-\phi_2^*\phi_3\\
-  \alpha\phi_1^*\phi_3&-\phi_2\phi_3^*&\alpha |\phi_3|^2+ \frac{ |\phi_1|^2}{\alpha}
\end{bmatrix} 
\end{array}
\end{eqnarray}
To prove that the matrix $\Lambda_{\alpha}\left(|\phi\ket\bra\phi|\right)$ is positive, we need to show that all of its principal minors are positive. The 1st order principal minors are the diagonal elements, which are positive for any $\alpha > 0$. The three 2nd order principal minors are 
\begin{eqnarray}
\begin{array}{ll}
M_1=  \frac{1}{\alpha + \frac{1}{\alpha}}\left|\begin{array}{ll}
\alpha (|\phi_1|^2+ |\phi_2|^2)~~ -\phi_1\phi_2^{*}\\
 -\phi_1^*\phi_2 ~~~~~~~~~~~~~\frac{|\phi_2|^2+|\phi_3|^2}{\alpha}
\end{array}\right| ,\\
\\
M_2= \frac{1}{\alpha + \frac{1}{\alpha}}\left|\begin{array}{ll}
\alpha (|\phi_1|^2+ |\phi_2|^2) ~~- \alpha\phi_1\phi_3^{*}\\

-  \alpha\phi_1^*\phi_3~~~~~~~~~~~~\alpha |\phi_3|^2+ \frac{ |\phi_1|^2}{\alpha}
\end{array}\right|,\\
\\
M_3= \frac{1}{\alpha + \frac{1}{\alpha}}\left|\begin{array}{ll}
\frac{|\phi_2|^2+|\phi_3|^2}{\alpha}~~&-\phi_2^*\phi_3\\
-\phi_2\phi_3^*~~~~~~~~~~~~\alpha |\phi_3|^2+ \frac{ |\phi_1|^2}{\alpha}
\end{array}\right|. 
\end{array}
\end{eqnarray}
We note that $M_1$ simplifies to $\frac{\alpha}{1+\alpha^{2}} \left( |\phi_1|^4 +|\phi_1|^2|\phi_3|^2 + |\phi_2|^2 |\phi_3|^2 \right) $. Therefore $M_1$ is non negative as $\alpha \in ( 0 , 1 ]$. Similarly, $M_2$ simplifies to $\frac{\alpha}{1+\alpha^{2}}  \left( \frac{|\phi_1|^4 + |\phi_1|^2|\phi_2|^2+|\phi_2|^2|\phi_3|^2 \alpha^{3}}{\alpha} \right) $ which is a non negative quantity,  and $M_3$ simplifies to $\frac{\alpha}{1+\alpha^{2}} \left( |\phi_3|^4 + \frac{|\phi_1|^2 (|\phi_2|^2+|\phi_3|^2)}{\alpha^{3}} \right)  $ which is again a non negative quantity as $\alpha \in ( 0 , 1 ]$.

The remaining principal minor is the determinant of the matrix $\Lambda_{\alpha}\left(|\phi\ket\bra\phi|\right)$, which is given by
\[
\begin{array}{ll}
D= \frac{\alpha^2}{1+\alpha^2}\left[|\phi_2|^2|\phi_3|^4+\frac{|\phi_3|^2|\phi_1|^4}{\alpha^2}+\frac{|\phi_1|^2|\phi_2|^4}{\alpha^2}\right]\\
- \frac{\alpha^2}{1+\alpha^2}\left[2|\phi_1|^2|\phi_2|^2|\phi_3|^2+2|\phi_1|^2Re(\phi_2^*\phi_3)^2)-\frac{1}{\alpha^2}|\phi_1|^2|\phi_2|^2|\phi_3|^2\right],\\
\\
\mbox{Since}~~ Re(\phi_2^*\phi_3)^2\leq |\phi_2|^2|\phi_3|^2, \forall~ \phi_2~~ \mbox{and}~~ \phi_3,~~ \mbox{we have}\\
\\
D\geq \frac{\alpha^2}{1+\alpha^2}\left[|\phi_2|^2|\phi_3|^4+\frac{|\phi_3|^2|\phi_1|^4}{\alpha^2}+\frac{|\phi_1|^2|\phi_2|^4}{\alpha^2}-(4-\frac{1}{\alpha^2})|\phi_1|^2|\phi_2|^2|\phi_3|^2\right],\\
\geq \frac{\alpha^2}{1+\alpha^2}\left[|\phi_2|^2|\phi_3|^4+|\phi_3|^2|\phi_1|^4+|\phi_1|^2|\phi_2|^4-3|\phi_1|^2|\phi_2|^2|\phi_3|^2\right],
\end{array}
\]
for all $\alpha\leq 1$. Here $Re(\cdot)$ means the real part of a complex number. It is straightforward to check that the quantity 
\[\left[|\phi_2|^2|\phi_3|^4+|\phi_3|^2|\phi_1|^4+|\phi_1|^2|\phi_2|^4-3|\phi_1|^2|\phi_2|^2|\phi_3|^2\right] \geq 0,\]
for all  $\phi_1, \phi_2,\phi_3$ with the constraint $|\phi_1|^2+|\phi_2|^2+|\phi_3|^2=1$. Therefore, the map $\Lambda_{\alpha}(\cdot)$ is positive for all $0 < \alpha\leq 1$. \qed

It is our aim to find whether the map $\Lambda_{\alpha}$ is useful to detect entangled states positive under partial transposition. For this purpose, we prove the following corollary. \\

\textbf{Corollary 1:} $\Lambda_{\alpha}$ contains a class of non completely positive indecomposable maps.

\proof To prove the corollary, we first have to show that the given positive map is is not completely positive. For this purpose, using Choi's theorem, it is sufficient to show that $\mathbb{I}\otimes\Lambda_{\alpha}(|\Phi\ket\bra\Phi|)$ is not positive. Here, $|\Phi\ket$ is the maximally entangled two qutrit state.

Let us consider the corresponding Choi matrix first. We take the maximally entangled state for two qutrit system as $ \vert \Phi \ket = \frac{1}{\sqrt{3}} \left( \vert 00 \ket + \vert 11 \ket + \vert 22 \ket \right) $ where,
\begin{eqnarray}
\vert 0 \ket = \begin{bmatrix}
1\\
0\\
0
\end{bmatrix}, ~~ \vert 1 \ket = \begin{bmatrix}
0\\
1\\
0
\end{bmatrix}, ~~ \vert 2 \ket = \begin{bmatrix}
0\\
0\\
1
\end{bmatrix}  
\end{eqnarray}

The one sided action of the map on the maximally entangled state gives rise to the Choi matrix,
\begin{eqnarray}
\begin{tiny}
\mathcal{C}_{\Lambda_{\alpha}}=\begin{bmatrix}
\frac{\alpha^2}{3 + 3 \alpha ^2}&0&0&0&- \frac{\alpha}{3+3 \alpha^2}&0&0&0&-\frac{\alpha^2}{3 + 3 \alpha ^2}\\
0&0&0&0&0&0&0&0&0\\
0&0&\frac{1}{3+3 \alpha^2}&0&0&0&0&0&0\\
0&0&0&\frac{\alpha^2}{3 + 3 \alpha ^2}&0&0&0&0&0\\
- \frac{\alpha}{3+3 \alpha^2}&0&0&0&\frac{1}{3+3 \alpha^2}&0&0&0&0\\
0&0&0&0&0&0&0&- \frac{\alpha}{3+3 \alpha^2}&0\\
0&0&0&0&0&0&0&0&0\\
0&0&0&0&0&- \frac{\alpha}{3+3 \alpha^2}&0&\frac{1}{3+3 \alpha^2}&0\\
-\frac{\alpha^2}{3 + 3 \alpha ^2}&0&0&0&0&0&0&0&\frac{\alpha^2}{3 + 3 \alpha ^2}
\end{bmatrix}
\end{tiny}
\end{eqnarray}
The least  eigenvalue of $\mathcal{C}_{\Lambda_{\alpha}}$ is $\lambda^{'}= \frac{1-\sqrt{1+4 \alpha^2}}{6+6 \alpha^2}$. We see that  it is a negative quantity within the above parameter range $\alpha \in ( 0 , 1 ]$. Hence, it is proven that the given map is not completely positive.

In the most straightforward way to prove that the map is indecomposable, we have to now show that it can detect at least one entangled state which is positive under partial transposition. Such a class of two qutrit entangled states \cite{StormerPPT} is the following
\begin{small}
\begin{eqnarray} \label{taux}
\tau_x=\frac{1}{3(1+x+x^{-1})}
\left(
\begin{array}{ccccccccc}
 1 & 0 & 0 & 0 & 1 & 0 & 0 & 0 & 1 \\
 0 & x & 0 & 0 & 0 & 0 & 0 & 0 & 0 \\
 0 & 0 & \frac{1}{x} & 0 & 0 & 0 & 0 & 0 & 0 \\
 0 & 0 & 0 & \frac{1}{x} & 0 & 0 & 0 & 0 & 0 \\
 1 & 0 & 0 & 0 & 1 & 0 & 0 & 0 & 1 \\
 0 & 0 & 0 & 0 & 0 & x & 0 & 0 & 0 \\
 0 & 0 & 0 & 0 & 0 & 0 & x & 0 & 0 \\
 0 & 0 & 0 & 0 & 0 & 0 & 0 & \frac{1}{x} & 0 \\
 1 & 0 & 0 & 0 & 1 & 0 & 0 & 0 & 1 \\
\end{array}
\right)
\end{eqnarray}
\end{small}
with $x$ being any non zero positive real number. Applying the proposed map, we have 

\[
\begin{tiny}
\begin{array}{ll}
\mathbb{I} \otimes \Lambda_{\alpha}(\tau_x)=\\ 

\frac{N}{3(1+x+x^{-1})}\begin{array}{ll}
\left(
\begin{array}{ccccccccc}
 \alpha  (x+1) & 0 & 0 & 0 & -1 & 0 & 0 & 0 & -\alpha  \\
 0 & \frac{x+\frac{1}{x}}{\alpha } & 0 & 0 & 0 & 0 & 0 & 0 & 0 \\
 0 & 0 & \frac{\alpha }{x}+\frac{1}{\alpha } & 0 & 0 & 0 & 0 & 0 & 0 \\
 0 & 0 & 0 & \alpha  \left(1+\frac{1}{x}\right) & 0 & 0 & 0 & 0 & 0 \\
 -1 & 0 & 0 & 0 & \frac{x+1}{\alpha } & 0 & 0 & 0 & 0 \\
 0 & 0 & 0 & 0 & 0 & x \alpha +\frac{1}{\alpha  x} & 0 & -1 & 0 \\
 0 & 0 & 0 & 0 & 0 & 0 & \alpha  \left(x+\frac{1}{x}\right) & 0 & 0 \\
 0 & 0 & 0 & 0 & 0 & -1 & 0 & \frac{1+\frac{1}{x}}{\alpha } & 0 \\
 -\alpha  & 0 & 0 & 0 & 0 & 0 & 0 & 0 & \frac{x}{\alpha }+\alpha  \\

\end{array}
\right)
\end{array}
\end{array}
\end{tiny}
\]
\\
Here, $N=1/(\alpha+1/\alpha)$ is the normalization factor. One of the principal minors of the matrix $\mathbb{I} \otimes \Lambda_{\alpha}(\tau_x)$ is given by 
\[
\begin{array}{ll}
D_{\tau_x}=N\left|\begin{array}{ll}
\alpha(1+x)~~-1~~-\alpha\\
-1~~~~~~~~~~~\frac{1+x}{\alpha}~~~~~~0\\
-\alpha~~~~~~~~~~~0~~\alpha+\frac{x}{\alpha}
\end{array}\right|\\
~~~~~~=N[x(2+x)(\alpha+\frac{x}{\alpha})-\alpha(1+x)]
\end{array}
\]
\\
The negativity of $D_{\tau_x}$ for a large range of parameters can be seen from Fig.1 
where we plot $D_{\tau_x}$ with respect to $x$ for some particular values of $\alpha$.
The following cases may be considered a examples.
Case 1: Considering $\alpha=\frac{1}{4}$, we see that $D_{\tau_x}<0$, if 
$x< 0.154.$
\\
Case 2: Considering $\alpha=\frac{1}{2}$, we see that $D_{\tau_x}<0$, if 
$x< 0.269.$
\\
Case 3: Let us now consider $\alpha=1$. In this case we can see that $D_{\tau_x}<0$, if 
$ x < \sqrt{2}-1 .$
\\
Therefore, it  the one parameter class of maps contains indecomposable positive maps. \qed \\

\begin{figure}[htb]
\resizebox{8cm}{5cm}{\includegraphics{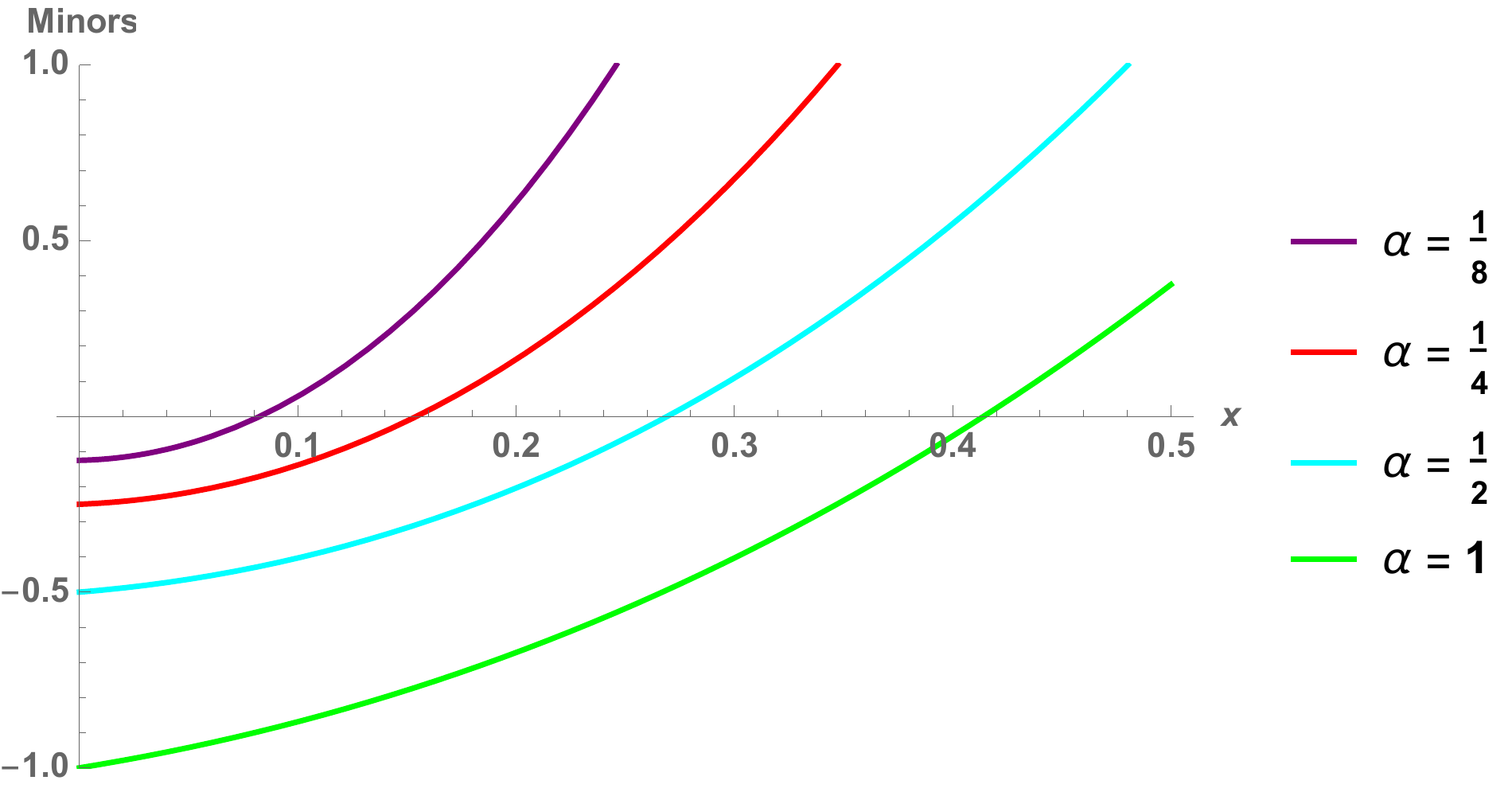}}
\caption{\footnotesize (Colour online) Variation of Minors $D_{\tau_x}$ with respect to map parameter $\alpha$}
\label{Minors} 
\end{figure}

\textbf{Remark:} In \cite{Yang16}, authors have shown that any indecomposable linear map $\Lambda : \mathbb{M}_3 \rightarrow \mathbb{M}_3 $ is atomic and hence in view of this fact the indecomposable maps in \textbf{Corollary 1} are also atomic.

 Let us now illustrate the dual map corresponding to the positive map introduced in the definition 1.

\noindent \textbf{Corollary 2:} The following map 
\beq\label{nmap}
\Lambda_{\alpha}^{\dagger} \left( X \right) = \frac{1}{\alpha + \frac{1}{\alpha}} \begin{bmatrix}
\alpha (x_{11}+ x_{33})& -x_{12}&- \alpha x_{13}\\
-x_{21}&\frac{x_{22}+x_{11}}{\alpha}&-x_{32}\\
- \alpha x_{31}&-x_{23}&\alpha x_{33}+ \frac{x_{22}}{\alpha}
\end{bmatrix}
\eeq
is also positive and indecomposable in the  range $0<\alpha\leq 1$.

\proof The proof of positivity follows similarly to that of \textbf{Theorem 1}. For the proof of indecomposability, we can construct the matrix $\mathbb{I}\otimes\Lambda_{\alpha}^{\dagger}(\tau_x)$, to find that it will have at least one negative eigenvalue for \[x>\frac{1}{\sqrt{2}-1}.\]

\qed

\section{Entanglement witness and Weak optimality}

Since positive maps are not physically realizable, it is our next goal to construct an entanglement witness class  \cite{guehne}, in order to set the experimental viability of our findings on a firm footing. Moreover,  we also prove that at least one of our constructed witnesses is weakly optimal.

Any positive but not completely positive map gives rise to an entanglement witness. For a given map $\Lambda_\Gamma$  its corresponding Choi matrix $\mathcal{C}_\Gamma$ serves as a witness for some entangled state. An entanglement witness $\mathcal{W}$ is said to be weakly optimal \cite{badzi} if there exists some pure product state $\vert \gamma \ket \otimes \vert \delta \ket$ such that 
 \[ \bra \gamma \vert \otimes \bra \delta \vert \mathcal{W} \vert \gamma \ket \otimes \vert \delta \ket = 0\]
 
An entanglement witness can always be constructed from a positive map. We give one such example in the following.

We know that for two positive semi-definite matrices $A$ and $B$, the identity $Tr[A.B]\geq 0$ always holds. Based on this fact, we have $Tr[|\Phi\ket\bra\Phi|\mathbb{I}\otimes\Lambda_{\alpha}^{\dagger}(\sigma)]\geq 0$, for all separable states $\sigma$ and for at least one entangled state, the trace identity will acquire negative value. Following the trace rule $Tr[C.D]=Tr[D.C]$ for any pair of matrices $C~\mbox{and}~D$, we get 

\[Tr[|\Phi\ket\bra\Phi|\mathbb{I}\otimes\Lambda_{\alpha}^{\dagger}(\rho)]=Tr[\mathbb{I}\otimes\Lambda_{\alpha}(|\Phi\ket\bra\Phi|)\rho]\geq 0,\] for any state $\rho$ of $3\times 3$ dimension. This can of course be extended to arbitrary $d\times 3$ dimensional systems.

We can thus consider the one parameter family of positive maps $\Lambda_{\alpha}$ with the corresponding Choi matrix $\mathcal{C}_{\Lambda_{\alpha}}=\mathbb{I}\otimes\Lambda_{\alpha}(|\Phi\ket\bra\Phi|)$, to be an one parameter family of entanglement witnesses. Now, for $\alpha = 1$, we can choose $ \vert \gamma \ket = \vert \delta \ket  = \frac{1}{3} \begin{bmatrix}
 1\\
 1\\
 1
 \end{bmatrix} $ such that 
 \[\bra \gamma \vert \otimes \bra \gamma \vert  \mathcal{C}_{\Lambda_{\alpha}} \vert \gamma \ket \otimes \vert \gamma \ket = 0\]      
 
 This weakly optimal witness can also be implemented locally. The witness can be expressed as a linear sum of two qutrit local observables.  We consider the $3 \times 3$ identity matrix 
$G_1= \begin{bmatrix}
1&0&0\\
0&1&0\\
0&0&1\\
\end{bmatrix}$,

along with 8 Gell-Mann matrices 


$G_2= \begin{bmatrix}
0&1&0\\
1&0&0\\
0&0&0\\
\end{bmatrix}$,
$G_3= \begin{bmatrix}
0&-i&0\\
i&0&0\\
0&0&0\\
\end{bmatrix}$,
$G_4= \begin{bmatrix}
1&0&0\\
0&-1&0\\
0&0&0\\
\end{bmatrix}$,

$G_5= \begin{bmatrix}
0&0&1\\
0&0&0\\
1&0&0\\
\end{bmatrix}$,
$G_6= \begin{bmatrix}
0&0&-i\\
0&0&0\\
1&0&0\\
\end{bmatrix}$,
$G_7= \begin{bmatrix}
0&0&0\\
0&0&1\\
0&1&0\\
\end{bmatrix}$,

$G_8= \begin{bmatrix}
0&0&0\\
0&0&-i\\
0&i&0\\
\end{bmatrix}$,
$G_9= \frac{1}{\sqrt{3}}\begin{bmatrix}
1&0&0\\
0&1&0\\
0&0&-2\\
\end{bmatrix}$

as 9 local  observables as they are Hermitian. We denote them as $G_i , i= 1 ..... 9$.
 
We note that for $\alpha=1$, 
 \beq\label{wit}
 \begin{array}{ll} 
 \mathcal{C}_{\Lambda_{1}} = \frac{1}{3} G_1 \otimes G_1 - \frac{1}{6}  G_2 \otimes G_2 + \frac{1}{6}  G_3 \otimes G_3 -\frac{1}{4 \sqrt{3}}  G_4 \otimes G_9\\
~~~~~~~~~~  -\frac{1}{6}  G_5 \otimes G_5+ \frac{1}{6}  G_6 \otimes G_6 -\frac{1}{6}  G_7 \otimes G_7
 -\frac{1}{6}  G_8 \otimes G_8 \\
 ~~~~~~~~~~+\frac{1}{4 \sqrt{3}}  G_9 \otimes G_4 
 +\frac{1}{12}  G_9 \otimes G_9. 
 \end{array}
 \nonumber
 \eeq
This witness is of course indecomposable, because the corresponding positive map is proven to be indecomposable. To further establish this fact, we apply the witness $ \mathcal{C}_{\Lambda_{1}}$ on $\tau_x$ (\ref{taux}) to find 
\beq
Tr\left[ \mathcal{C}_{\Lambda_{1}}\tau_x\right]=\frac{3-x}{18\left(x^2+x+1\right)}.
\eeq
It is thus clear that the witness  detects entanglement of the two-qutrit entangled state  $\tau_x$ for $x>3$.

\section{Structural physical approximation and a new class of states with PPT entanglement}

The structural physical approximation (SPA)\citep{spa1,spa2} of a positive map is a convex mixture of a depolarizing map with the given map, so that the resulting map is complete positive.  A map $\Lambda_{dep} : \mathbb{M}_d \rightarrow \mathbb{M}_d$ is said to be depolarizing if $\Lambda_{dep} \left( X \right) = \frac{Tr\left( X \right) }{d} \mathbb{I}$ for $X \in \mathbb{M}_d $.  Mathematically, SPA maps are the points of intersection of the line joining the given map with the depolarizing map and the set of all complete positive maps. Operationally, SPA of a positive map is obtained by adding some disturbance to the positive map. 

An algorithm to find the optimal SPA map for a given positive map has  been prescribed in \citep{spa1}.  We shall now formulate the SPA corresponding to the one parameter family of maps $\Lambda_{\alpha}$ and show that it gives rise to a class of PPT entangled states.
We have earlier considered the Choi matrix $\mathcal{C}_{\Lambda_{\alpha}}$ of the family of maps and found
the least  eigenvalue of $\mathcal{C}_{\Lambda_{\alpha}}$ to be $\lambda^{'}= \frac{1-\sqrt{1+4 \alpha^2}}{6+6 \alpha^2}$ when we take $\alpha \in ( 0 , 1 ]$. It is a negative quantity within the above parameter range. Defining $\lambda = max \left[ 0 , -\lambda^{'} \right] $, and
following the prescription of \citep{spa1}, the optimal SPA map corresponding to the map $\Lambda_{\alpha}$ is given by
\begin{eqnarray}
\Lambda_{\alpha}^{opt} = p^* \Lambda_{dep} + (1-p^*) \Lambda_{\alpha}^{'} \nonumber
\end{eqnarray}
where $p^* = \frac{\lambda d d^{'} \beta_{\Lambda_{\alpha}}^{-1}}{\lambda d d^{'} \beta_{\Lambda}^{-1}+ 1}$,  $\Lambda_{dep} = \frac{Tr\left( . \right) }{d^{'}} \mathbb{I}$, and $\Lambda^{'}= \beta_{\Lambda}^{-1} \Lambda$  is the re-scaling of the original map. Here $d = d^{'} = 3$, the input and output dimension of the map $\Lambda_{\alpha}$, and as a consequence of trace preservation, the value of $ \beta_{\Lambda_{\alpha}} = 1$.
Therefore, the optimal SPA map $\Lambda_{\alpha}^{opt} : \mathbb{M}_3 \rightarrow \mathbb{M}_3 $  is given by
\\
\\
\begin{widetext}
\begin{eqnarray}
\Lambda_{\alpha}^{opt} \left( X \right) = \left[
\begin{small}
\begin{array}{ccc}
 \frac{x_{33} \left(\sqrt{4 \alpha ^2+1}-1\right)+x_{11} \left(2 \alpha ^2+\sqrt{4 \alpha ^2+1}-1\right)+x_{22} \left(2 \alpha ^2+\sqrt{4 \alpha ^2+1}-1\right)}{2 \alpha ^2+3 \sqrt{4 \alpha ^2+1}-1} & \frac{2 x_{12} \alpha }{-2 \alpha ^2-3 \sqrt{4 \alpha ^2+1}+1} & \frac{2 x_{13} \alpha ^2}{-2 \alpha ^2-3 \sqrt{4 \alpha ^2+1}+1} \\
 \frac{2 x_{21} \alpha }{-2 \alpha ^2-3 \sqrt{4 \alpha ^2+1}+1} & \frac{x_{11} \left(\sqrt{4 \alpha ^2+1}-1\right)+(x_{22}+x_{33}) \left(\sqrt{4 \alpha ^2+1}+1\right)}{2 \alpha ^2+3 \sqrt{4 \alpha ^2+1}-1} & \frac{2 x_{32} \alpha }{-2 \alpha ^2-3 \sqrt{4 \alpha ^2+1}+1} \\
 \frac{2 x_{31} \alpha ^2}{-2 \alpha ^2-3 \sqrt{4 \alpha ^2+1}+1} & \frac{2 x_{23} \alpha }{-2 \alpha ^2-3 \sqrt{4 \alpha ^2+1}+1} & \frac{-x_{22}+x_{33} \left(2 \alpha ^2-1\right)+x_{11} \left(\sqrt{4 \alpha ^2+1}+1\right)+(x_{22}+x_{33}) \sqrt{4 \alpha ^2+1}}{2 \alpha ^2+3 \sqrt{4 \alpha ^2+1}-1} \\
\end{array}
\end{small}
\right ]
\end{eqnarray}
\end{widetext}
We note that the SPA map is also trace preserving. To check whether the SPA is completely positive, we compute the  corresponding Choi matrix. The Choi matrix is found to be
\begin{widetext}
\begin{eqnarray}
\mathcal{C}_{\Lambda_{\alpha}^{opt}} = \begin{tiny} \left[
\begin{array}{ccccccccc}
 \frac{2 \alpha ^2+\sqrt{4 \alpha ^2+1}-1}{6 \alpha ^2+9 \sqrt{4 \alpha ^2+1}-3} & 0 & 0 & 0 & \frac{2 \alpha }{-6 \alpha ^2-9 \sqrt{4 \alpha ^2+1}+3} & 0 & 0 & 0 & \frac{2 \alpha ^2}{-6 \alpha ^2-9 \sqrt{4 \alpha ^2+1}+3} \\
 0 & \frac{\sqrt{4 \alpha ^2+1}-1}{6 \alpha ^2+9 \sqrt{4 \alpha ^2+1}-3} & 0 & 0 & 0 & 0 & 0 & 0 & 0 \\
 0 & 0 & \frac{\sqrt{4 \alpha ^2+1}+1}{6 \alpha ^2+9 \sqrt{4 \alpha ^2+1}-3} & 0 & 0 & 0 & 0 & 0 & 0 \\
 0 & 0 & 0 & \frac{2 \alpha ^2+\sqrt{4 \alpha ^2+1}-1}{6 \alpha ^2+9 \sqrt{4 \alpha ^2+1}-3} & 0 & 0 & 0 & 0 & 0 \\
 \frac{2 \alpha }{-6 \alpha ^2-9 \sqrt{4 \alpha ^2+1}+3} & 0 & 0 & 0 & \frac{\sqrt{4 \alpha ^2+1}+1}{6 \alpha ^2+9 \sqrt{4 \alpha ^2+1}-3} & 0 & 0 & 0 & 0 \\
 0 & 0 & 0 & 0 & 0 & \frac{\sqrt{4 \alpha ^2+1}-1}{6 \alpha ^2+9 \sqrt{4 \alpha ^2+1}-3} & 0 & \frac{2 \alpha }{-6 \alpha ^2-9 \sqrt{4 \alpha ^2+1}+3} & 0 \\
 0 & 0 & 0 & 0 & 0 & 0 & \frac{\sqrt{4 \alpha ^2+1}-1}{6 \alpha ^2+9 \sqrt{4 \alpha ^2+1}-3} & 0 & 0 \\
 0 & 0 & 0 & 0 & 0 & \frac{2 \alpha }{-6 \alpha ^2-9 \sqrt{4 \alpha ^2+1}+3} & 0 & \frac{\sqrt{4 \alpha ^2+1}+1}{6 \alpha ^2+9 \sqrt{4 \alpha ^2+1}-3} & 0 \\
 \frac{2 \alpha ^2}{-6 \alpha ^2-9 \sqrt{4 \alpha ^2+1}+3} & 0 & 0 & 0 & 0 & 0 & 0 & 0 & \frac{2 \alpha ^2+\sqrt{4 \alpha ^2+1}-1}{6 \alpha ^2+9 \sqrt{4 \alpha ^2+1}-3} \\
\end{array}
\right]
\end{tiny}
\end{eqnarray}
\end{widetext}

We next compute the eigenvalues of the $9 \times 9$ Choi matrix $\mathcal{C}_{\Lambda_{\alpha}^{opt}}$  and observe that  the Choi matrix is positive semi-definite for the whole range of $\alpha$ as all of its eigenvalues are non negative for the whole range of $\alpha$. This signifies that the SPA map is complete positive. Moreover, it is to be noted that the Choi matrix is a valid density matrix as  $\mathcal{C}_{\Lambda_{\alpha}^{opt}}$ is Hermitian, positive semi-definite and of trace $1$ for $ \alpha \in (0, 1]$. Hence, we obtain a one parameter family of two qutrit states.

The partial transposition of $\mathcal{C}_{\Lambda_{\alpha}^{opt}}$ is given by
\begin{widetext}
\begin{small}
$\mathcal{C}^{T}_{\Lambda_{\alpha}^{opt}}$ = $\left(
\begin{array}{ccccccccc}
 \frac{2 \alpha ^2+\sqrt{4 \alpha ^2+1}-1}{6 \alpha ^2+9 \sqrt{4 \alpha ^2+1}-3} & 0 & 0 & 0 & 0 & 0 & 0 & 0 & 0 \\
 0 & \frac{\sqrt{4 \alpha ^2+1}-1}{6 \alpha ^2+9 \sqrt{4 \alpha ^2+1}-3} & 0 & \frac{2 \alpha }{-6 \alpha ^2-9 \sqrt{4 \alpha ^2+1}+3} & 0 & 0 & 0 & 0 & 0 \\
 0 & 0 & \frac{\sqrt{4 \alpha ^2+1}+1}{6 \alpha ^2+9 \sqrt{4 \alpha ^2+1}-3} & 0 & 0 & 0 & \frac{2 \alpha ^2}{-6 \alpha ^2-9 \sqrt{4 \alpha ^2+1}+3} & 0 & 0 \\
 0 & \frac{2 \alpha }{-6 \alpha ^2-9 \sqrt{4 \alpha ^2+1}+3} & 0 & \frac{2 \alpha ^2+\sqrt{4 \alpha ^2+1}-1}{6 \alpha ^2+9 \sqrt{4 \alpha ^2+1}-3} & 0 & 0 & 0 & 0 & 0 \\
 0 & 0 & 0 & 0 & \frac{\sqrt{4 \alpha ^2+1}+1}{6 \alpha ^2+9 \sqrt{4 \alpha ^2+1}-3} & 0 & 0 & 0 & \frac{2 \alpha }{-6 \alpha ^2-9 \sqrt{4 \alpha ^2+1}+3} \\
 0 & 0 & 0 & 0 & 0 & \frac{\sqrt{4 \alpha ^2+1}-1}{6 \alpha ^2+9 \sqrt{4 \alpha ^2+1}-3} & 0 & 0 & 0 \\
 0 & 0 & \frac{2 \alpha ^2}{-6 \alpha ^2-9 \sqrt{4 \alpha ^2+1}+3} & 0 & 0 & 0 & \frac{\sqrt{4 \alpha ^2+1}-1}{6 \alpha ^2+9 \sqrt{4 \alpha ^2+1}-3} & 0 & 0 \\
 0 & 0 & 0 & 0 & 0 & 0 & 0 & \frac{\sqrt{4 \alpha ^2+1}+1}{6 \alpha ^2+9 \sqrt{4 \alpha ^2+1}-3} & 0 \\
 0 & 0 & 0 & 0 & \frac{2 \alpha }{-6 \alpha ^2-9 \sqrt{4 \alpha ^2+1}+3} & 0 & 0 & 0 & \frac{2 \alpha ^2+\sqrt{4 \alpha ^2+1}-1}{6 \alpha ^2+9 \sqrt{4 \alpha ^2+1}-3} \\
\end{array}
\right)$
\end{small}
\end{widetext}
We  compute its eigenvalues and plot them with respect to $\alpha$ in Figure 2. We note that among the nine eigenvalues, one eigenvalue is negative in the interval $\alpha \in (0, 1)$, and other eight eigenvalues are all positive for $\alpha \in (0, 1)$ . 
Hence, the class of states for this interval of values of $\alpha$ is NPT, and therefore, it is entangled. Interestingly, for the  parameter value $\alpha= 1$, all the eigenvalues of the partially transposed matrix are non negative, and hence, the state $\mathcal{C}_{\Lambda_{\alpha}^{opt}}$ is PPT for $\alpha = 1$. 
\begin{figure}[htb]
\resizebox{8cm}{5cm}{\includegraphics{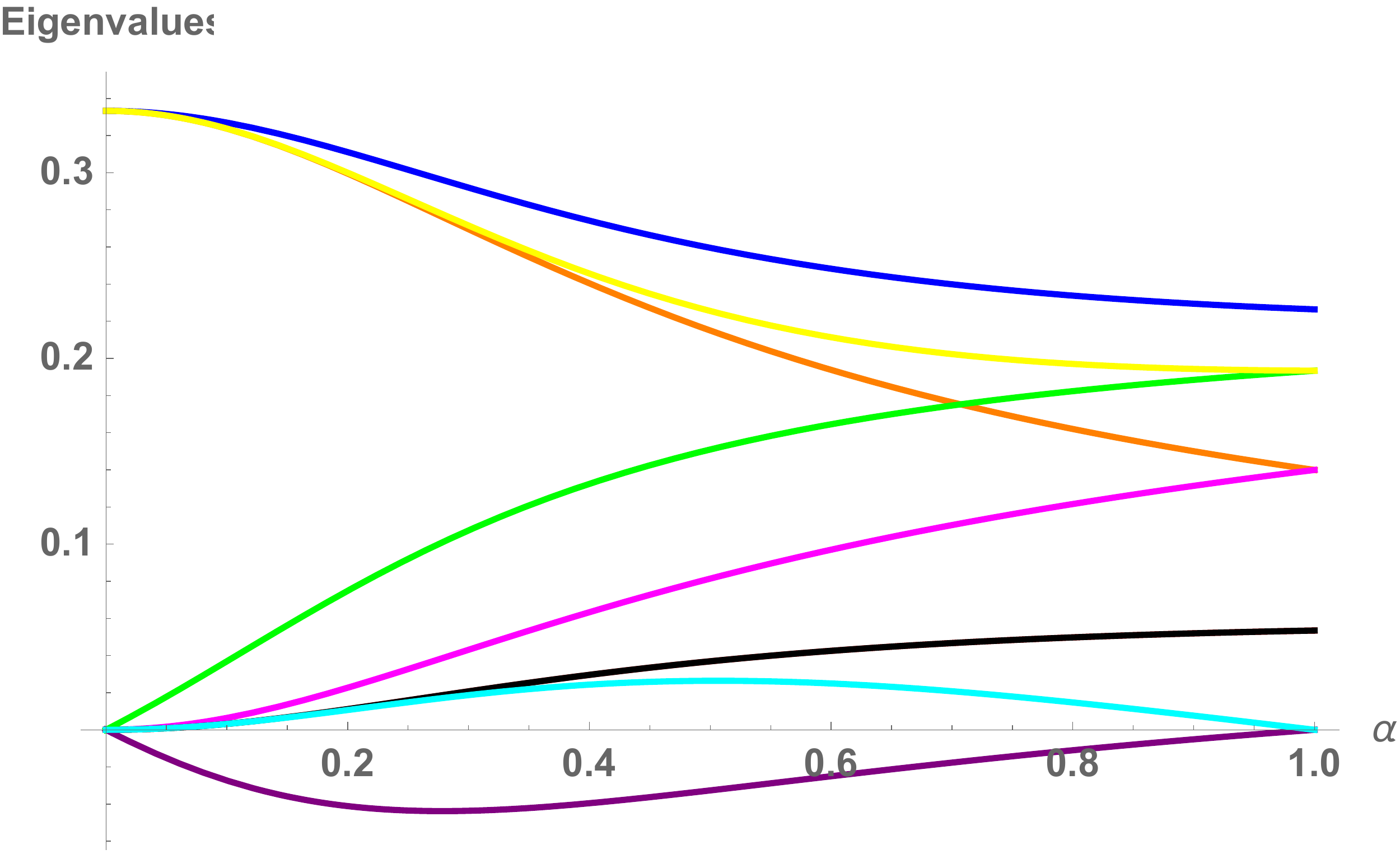}}
\caption{(Colour online) The eigenvalues of the  partially transposed matrix $\mathcal{C}^{T}_{\Lambda_{\alpha}^{opt}}$ are plotted with respect to $\alpha$. }
\label{fig3}
\end{figure}


We are interested to find whether the state $\mathcal{C}_{\Lambda_{1}^{opt}}$ is entangled. It may be noted that since we are dealing with a two-qutrit system, the partial transposition criterion is no longer sufficient for entanglement detection. 
We hence adopt the covariance matrix criterion \cite{GHGE_07, GGHE_08} for entanglement detection.  We can consider the 3 by 3 identity matrix along with the 8 Gell-Mann matrices as orthogonal local observables as they are orthogonal and Hermitian. 
We take the Choi state $\mathcal{C}_{\Lambda_{\alpha}^{opt}}$  and obtain its two reduced density matrices $\mathcal{C}_{\Lambda_{\alpha}^{opt}}^{A}$ and  $\mathcal{C}_{\Lambda_{\alpha}^{opt}}^{B}$ respectively. The covariance matrix criterion \cite{GHGE_07, GGHE_08} states that for separable states,
\begin{eqnarray}
\label{cmatrix}
\vert\vert C \vert\vert_{1} \leq \sqrt{\left( 1- Tr \left[ \left(\mathcal{C}_{\Lambda_{\alpha}^{opt}}^{A} \right) ^{2}\right]  \right) \left( 1- Tr \left[ \left( \mathcal{C}_{\Lambda_{\alpha}^{opt}}^{B}\right) ^{2}\right] \right)} 
\end{eqnarray}
where $\vert\vert. \vert\vert_{1}$ stands for the trace norm and the components of the $C$ matrix are given by 
\begin{eqnarray}
C_{ij}= \bra H_{i}^{A} \otimes H_{j}^{B} \ket - \bra H_{i}^{A} \ket \bra H_{j}^{B} \ket
\label{Cmat}
\end{eqnarray} \\\\
and  $H_{i}^{A}$ and  $H_{j}^{B}$ denote local orthogonal observables on two sides.  We 
evaluate the C matrix using the  state  $\mathcal{C}_{\Lambda_{\alpha}^{opt}}$ and its reduced density matrices,  and find that the LHS  of  Eq.(\ref{cmatrix}) is strictly greater than the RHS for the values of $\alpha$ in $ \left(0 , 1 \right] $. The result has been illustrated in Fig. 3. This certifies the presence of entanglement in the state $\mathcal{C}_{\Lambda_{\alpha}^{opt}}$ for $\alpha \in  \left(0 , 1 \right] $. Therefore, the state corresponding to the value of the parameter $\alpha = 1  $, \textit{i.e}., $\mathcal{C}_{\Lambda_{1}^{opt}}$ is PPT-entangled. So, the SPA map $\Lambda_{\alpha}^{opt}$ can generate a  two-qutrit PPT entangled state $\mathcal{C}_{\Lambda_{1}^{opt}}$.  

\begin{figure}[htb]
\resizebox{8cm}{5cm}{\includegraphics{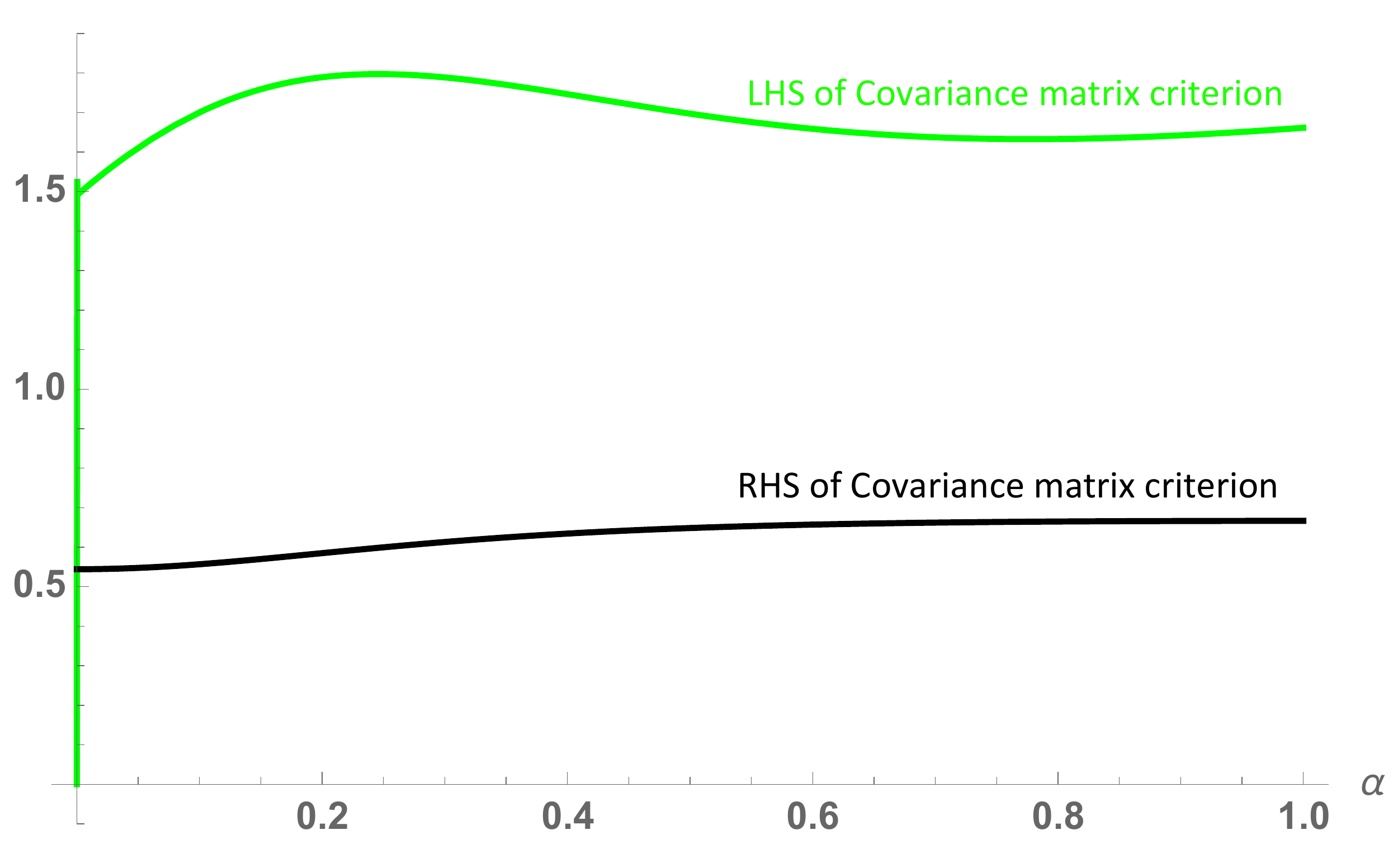}}
\caption{(Colour online) LHS and RHS of Eq.(\ref{cmatrix}) are plotted versus $\alpha$. }
\label{fig3}
\end{figure}
It may be further noted that the state $\mathcal{C}_{\Lambda_{\alpha}^{opt}}$ is NPT for the values of the parameter $\alpha $ in $\left( 0 , 1 \right) $ and therefore, it is entangled. 
  For the  parameter value  $\alpha = 1$ the  state is PPT and its entanglement can be detected via the covariance matrix criterion. Moreover, from figure 3 it is clear that the covariance matrix criterion can also detect the the entanglement in the range $\left( 0 , 1 \right) $ where the state is NPT.

Finally, let us check whether the PPT entangled state $\mathcal{C}_{\Lambda_{1}^{opt}}$ can be detected by some other existing positive maps. 
PPT entangled states are considered as a weak form of entanglement that is usually very hard to detect. As discussed earlier,  indecomposable maps are necessary to detect PPT entangled states. The celebrated Choi map $ \phi_{choi} : \mathbb{M}_3 \rightarrow \mathbb{M}_3 $  \citep{Choi75}, one of the first examples of indecomposable maps in the literature,  is defined as
\begin{eqnarray}
 \phi_{Choi} \left( X \right) =  \begin{bmatrix}
x_{11}+ x_{22} & -x_{12}& - x_{13}\\
-x_{21}&x_{22}+x_{11}&-x_{23}\\
-  x_{31}&-x_{32}& x_{33}+x_{22}
\end{bmatrix} 
\end{eqnarray}
where
\begin{eqnarray}
 X= \begin{bmatrix}
x_{11}&x_{12}&x_{13}\\
x_{21}&x_{22}&x_{23}\\
x_{31}&x_{32}&x_{33}&
\end{bmatrix} \in \mathbb{M}_3. \nonumber
\end{eqnarray} 
A positive map $\Lambda$ is said to detect an entangled state $\kappa$ if and only if $\mathbb{I} \otimes \Lambda \left( \kappa \right) <\Theta $, where $\Theta$ stands for zero operator. It can be checked that $\mathbb{I} \otimes \phi_{Choi} \left( \mathcal{C}_{\Lambda_{\alpha}^{opt}} \right) \geqslant \Theta ~~ \forall \alpha \in ( 0 , 1 ]$.
  Hence, the Choi map cannot detect the above PPT entangled state. 
 
 Recently, Miller and Olkiewicz \citep{ Miller15} introduced another indecomposable map  $\phi_{MO}$ on $\mathbb{M}_3$ given by,
\begin{eqnarray}
 \phi_{MO} \left( X \right) =  \begin{bmatrix}
\dfrac{1}{2} (x_{11}+ x_{22}) & 0& \frac{1}{\sqrt{2}} x_{13}\\
0&\dfrac{1}{2} (x_{11}+ x_{22})&\frac{1}{\sqrt{2}}x_{32}\\
\frac{1}{\sqrt{2}}  x_{31}&\frac{1}{\sqrt{2}}x_{23}& x_{33}
\end{bmatrix} 
\end{eqnarray}
where
\begin{eqnarray}
X= \begin{bmatrix}
x_{11}&x_{12}&x_{13}\\
x_{21}&x_{22}&x_{23}\\
x_{31}&x_{32}&x_{33}&
\end{bmatrix} \in \mathbb{M}_3. \nonumber
\end{eqnarray} 
It can be again checked that $\mathbb{I} \otimes \phi_{MO} \left( \mathcal{C}_{\Lambda_{\alpha}^{opt}} \right) \geqslant \Theta ~~ \forall \alpha \in ( 0 , 1 ]$.
 Hence the above map also cannot detect the PPT entangled state $\mathcal{C}_{\Lambda_{\alpha}^{opt}} $.

\section{conclusions}

Bound entangled states are hard to find and detect. In this work, we have constructed a one parameter family of indecomposable positive maps in three dimensional Hilbert space. These maps are shown to detect entanglement in a certain class of
two-qutrit PPT entangled states. Through our proposed non-completely positive indecomposable map we are additionally able to find a new class of entangled states
among which there exists a PPT entangled state.
We have further constructed  a weak optimal entanglement witness from one of these maps and have given its representation in terms of local observables. This presents the way to physically implement this  witness towards
detection of the two-qutrit bound entangled state. 

Moreover, we have also considered the structural physical approximation \cite{spa1, spa2}of the proposed positive map. This leads to a large class of NPT entangled states, but more interestingly, we have found a unique  bound entangled state which cannot be detected by various other well-known non-completely positive maps  \citep{Choi75,Miller15}. PPT entangled states have been constructed earlier from indecomposable positive maps using geometrical 
methods \cite{Ha03, HA04}.  In the present analysis we have devised a new procedure of contructing PPT entangled states employing the structural physical approximation. To conclude, this work 
motivates further investigations of positive maps and their applications in entanglement theory in higher dimensions.

{\it Acknowledgements:} NG would like to acknowledge support from the Research Initiation Grant BITS/GAU/RIG/2019/H0680
of BITS-Pilani, Hyderabad. ASM acknowledges support from the 
project DST/ICPS/QuEST/Q98 from the Department of Science and Technology, India. BB acknowledges the support from DST INSPIRE programme.

\bibliography{PPT_ref.bib}

\end{document}